\documentstyle[12pt]{article}

\def\simlt{\lower.5ex\hbox{$\; \buildrel < \over \sim \;$}}
\def\simgt{\lower.5ex\hbox{$\; \buildrel > \over \sim \;$}}

\newcommand{\sh}{\mbox{sinh}}

\newcommand{\puis}[1]{$^{#1}$}
\newcommand{\beq}{\begin{equation}}
\newcommand{\eeq}{\end{equation}}
\newcommand{\bea}{\begin{eqnarray}}
\newcommand{\ena}{\end{eqnarray}}
\newcommand{\EP}{\mbox{$E_{p}$}}

\newcommand{\EPB}{\mbox{$E_{\bar{p}}$}}

\newcommand{\NPB}{\mbox{$N_{\bar{p}}$}}
\newcommand{\YIELD}{\mbox{$Y_{\bar{p}}(E_{p} \to E_{\bar{p}})$}}

\begin{document}
\begin{titlepage}
\rightline{CERN-TH/96-114}
\rightline{ENSLAPP-A-550/96}

\vskip 0.5cm
\begin{center}
{\Large \bf Galactic diffusion and the antiprotron signal of
supersymmetric dark matter}
\vskip 0.5cm
Pascal Chardonnet$^{\rm a,b}$,
Giulio Mignola$^{\rm c}$,
Pierre Salati$^{\rm a,b,d}$ and
Richard Taillet$^{\rm a,b}$
\end{center}

\vskip 0.5cm
\begin{flushleft}
{\it
a) Theoretical Physics Group ENSLAPP, BP110, F-74941 Annecy-le-Vieux
Cedex, France.\\
b) Universit\'e de Savoie, BP1104 73011 Chamb\'ery Cedex,
France.\\
c) Theoretical Physics Division, CERN, CH-1211 Geneva 23,
Switzerland and I.N.F.N, Sez. di Torino, via P. Giuria 1, I-10125 Turin, Italy.
\\ 
d) Institut Universitaire de France.}
\end{flushleft}

\vskip 0.2cm
\begin{abstract}
The leaky box model is now ruled out by measurements of a cosmic ray
gradient throughout the galactic disk. It needs to be replaced by a more
refined treatment which takes into account the diffusion of cosmic rays
in the magnetic fields of the Galaxy. We have estimated the flux of
antiprotons on the Earth in the framework of a two-zone diffusion model.
Those species are created by the spallation reactions of high-energy
nuclei with the interstellar gas. Another potential source of antiprotons
is the annihilation of supersymmetric particles in the dark halo that
surrounds our Galaxy. In this letter, we investigate both processes.
Special emphasis is given to the antiproton signature of supersymmetric
dark matter. The corresponding signal exceeds the conventional
spallation flux below 300 MeV, a domain that will
be thoroughly explored by the Antimatter Spectrometer experiment.
The propagation of the antiprotons produced in the remote regions of
the halo back to the Earth plays a crucial role. Depending on the
energy, the leaky box estimates are wrong by a factor varying from 0.5 up
to 3.
\end{abstract}
\vspace{.2truecm}
CERN--TH/96--114 \hfill \break
June 1996

\end{titlepage}

\section{Introduction}

The flux of antiprotons on the Earth is about to be measured with
unprecedented accuracy by the Antimatter Spectrometer experiment
(AMS) \cite{AMS}.
A detector will be shipped in orbit on the space shuttle.
Cosmic rays will be directly detected in space. Unlike in balloon-borne
experiments, measurements will not be spoilt by the atmosphere.
Even in the stratosphere, the column density of air is still of the order
of a few g$\cdot$cm\puis{-2}. This value is still an order of magnitude above
the average of the column density of interstellar gas toward the region
of the galactic centre. The AMS mission will be mostly sensitive
to the energy range extending between 100 MeV and a few GeV, in
a domain so far barely explored and where previous measurements are
inconsistent with one another. The proton and antiproton fluxes
will give precious indications on the propagation of cosmic rays
at low energy.

As detailed in section~2, antiprotons are predominantly produced
by the interactions of primary cosmic rays, mostly protons, with the
interstellar gas that pervades the thin matter disk of our Galaxy.
Quite exciting is the possible presence of an additional antiproton
flux at low energy, which would signal the presence of massive
neutral particles around our Galaxy. The Milky Way is embedded
inside a halo of invisible material whose nature has not yet been
determined. That dark matter could be made of relics from the
big-bang. In that case, supersymmetric species are a favoured option
and AMS should be sensitive to the antiprotons produced by their mutual
annihilations.

Our understanding of cosmic ray propagation throughout the Galaxy
has recently improved. In the old leaky box scheme,
the Galaxy is pictured as one single region where 
production, interactions and diffusion of cosmic rays take place
homogeneously. Cosmic rays are stored in a vast domain, from which
they escape after being confined for some time. According to the leaky box
model, they are uniformly distributed. This last feature is not quite correct.
Measurements of the $\gamma$-ray diffuse emission of the Galaxy
actually point toward a gradient of the cosmic ray density along the
galactic ridge \cite{BLOEMEN_GAMMA}. The proton flux varies
with a galactocentric radius $r$.
Recently, the $\gamma$-ray diffuse radiation of the Perseus arm region
has been accurately determined by the EGRET experiment on board
the Compton Gamma Ray Observatory. The $\gamma$-ray emissivity
of interstellar hydrogen atoms immersed in the cosmic radiation drops
by a factor of $\sim$ 1.7 between the solar circle ($r \sim 9$ kpc)
and the Perseus arm ($r \sim 13$ kpc) \cite{GRO_CEPHEUS}.
The naive leaky box model needs actually to be replaced by a
more refined treatment of cosmic ray
propagation.

In section~3, a two-zone diffusion model is presented.
Cosmic ray primaries are produced inside a thin matter disk in which
they propagate and interact with the interstellar gas. Those spallations
generate antiprotons, among a variety of secondary nuclei. Cosmic rays
diffuse in the chaotic magnetic fields carried by the galactic stellar winds.
That diffusion takes place in an extended region, which encapsulates
the previous thin disk and extends a few kiloparsecs away from
the galactic plane. In this letter, we focus on the antiproton flux on
the Earth, which is due to the annihilation of supersymmetric particles in the
galactic halo. The corresponding antiproton yield is also derived
in that section, in the framework of diffusion.

Finally, in section~4, we discuss our results. Massive neutral species
annihilate in remote regions of the Galaxy, and antiprotons propagate back
to the solar system. Their diffusion from the outer parts of the
halo all the way toward the Earth is a crucial mechanism, which
cannot be accounted for by the naive leaky box model. However,
the main trend is preserved: the supersymmetric antiprotons
show up at low energy. Above $\sim$ 1 GeV, they are swamped
in the flux of spallation antiprotons. The strongest
supersymmetric signal is generated either by a pure gaugino or
a pure higgsino. The intermediate case, where the supersymmetric
species is a mixture of a gaugino and a higgsino, will be barely detectable,
unless the abundance of that particle in the galactic halo is larger
than its cosmological average. We also compare our predictions to
the leaky box estimates. A correct treatment of the cosmic ray
diffusion throughout the Galaxy is important. In the
leaky box model, the antiproton flux is underestimated at low
energy by a factor of 2 and overestimated at high energy by a factor
of $\sim$ 3. We have also derived the confinement time of the
antiprotons inside the galactic halo and disk as a function of energy.
Such a result may be directly incorporated in a leaky box model to
yield a not really bad estimate of the antiproton flux. We feel that
it may be helpful to colleagues who would like to rapidly derive a
fair value of the antiproton signal without getting into the intricacies
of our more involved treatment.
\newpage

\section{ \bf The sources of antiprotons }


Antiprotons are naturally produced by the nuclear
interactions of cosmic ray particles with the interstellar medium.
In order to estimate the energy spectrum of the antiprotons created
during a single collision, we have used the Lund Monte Carlo simulations PHYTIA
and JETSET \cite{BENGTSSON}. Since both cosmic rays and the
interstellar medium are mostly composed of protons, the fundamental
process is the collision of a high-energy proton with a proton at rest in
the interstellar medium. This reaction produces hadrons, mostly pions
whose radiative decays are the principal source of the galactic diffuse
gamma-ray emission. During the fragmentation process, when the inelastic
collision produces coloured strings, resonances such as
$\Lambda^{0}$, $\Delta^{-}$, $\Delta^{--}$ or $\Sigma^-$ are formed
and subsequently decay into antiprotons. Spallation reactions also produce
antineutrons, which in turn decay into antiprotons. The threshold for the
antiproton production mechanism
$p + p \rightarrow 3p + \overline{p}$
is 7 GeV. The energy $\EP$ of the incoming cosmic ray proton has been
varied from that threshold up to 2.7 TeV. For each value of $\EP$, a
million events have been generated to yield the antiproton multiplicity
per collision
\beq
\YIELD \; = \;
{\displaystyle \frac{d\NPB}{d\EPB}} \;\; ,
\label{yield}
\eeq
in the antiproton momentum range extending from 100 MeV to
100 GeV.

One of the best motivated particle candidates for cold dark matter
is provided by supersymmetry \cite{EHNOS}.
The lightest supersymmetric particle (LSP) is stable, provided R-parity is
conserved. In a large region of the supersymmetric parameter space,
that species is the lightest neutralino ($\chi$), defined as the lowest-mass
superposition of the gaugino ($\tilde{\gamma},\tilde{Z}$) and the higgsino
($\tilde{H_1},\tilde{H_2}$) fields~:
\beq
\chi \; = \; a_1 \tilde{\gamma} + a_2 \tilde{Z} + a_3 \tilde{H_1} +
a_4 \tilde{H_2} \;\; .
\eeq
The possibility to detect the presence of neutralinos in our galactic halo
has been studied extensively in many different ways \cite{GKJ}.
Direct detection of the nuclear recoil induced by neutralino--nucleus
elastic scattering is, both theoretically \cite{th_dir} and experimentally
\cite{exp_dir}, under investigation. The indirect detection of high-energy
neutrinos coming from the centre of the Earth or the Sun \cite{nu}, or
the indirect detection of exotic components in the primary cosmic rays,
such as $\gamma$, $e^+$ and $\bar{p}$, have been studied in detail
\cite{antip_other,antip_noi}. Here we examine the latter possibility, and
concentrate on the diffusion of the cosmic antiprotons $\bar{p}$ from the
remote regions of their production back to the inner Galaxy and the Earth.

The differential rate (per unit volume and per second) for the production
of antiprotons from $\chi$--$\chi$ annihilations is given by
\beq
q_{\bar{p}}^{\rm susy}(E_{\bar{p}}) \; = \;
\langle\sigma v\rangle f(E_{\bar p})
\left\{ \frac{\rho_{\chi}}{m_{\chi}}\right\}^2 \;\; ,
\label{susy_source}
\eeq
in which $E_{\bar{p}}$ denotes the antiproton energy, $\sigma$ is the
$\chi$--$\chi$ annihilation cross section and $v$  is the neutralino velocity
in the galactic halo. The neutralino density $\rho_{\chi}$ varies with
its position. For a single annihilation, the antiproton energy spectrum is
\beq
f(E_{\bar p}) \; \equiv \;
\left\{ {1 \over \sigma} \right\} \;
\left\{
{{d \sigma (\chi \chi \rightarrow \bar{p} + X)} \over {dE_{\bar{p}}}}
\right\} \; = \; {\displaystyle \sum_{F,f}} \;
B^{(F)}_{\chi f} \;\;\;
\left( {dN^{f}_{\bar{p}}\over dE_{\bar{p}}} \right) \;\; ,
\eeq
where $F$ describes the $\chi$--$\chi$ annihilation final state and
$B^{(F)}_{\chi f}$ is the branching ratio into the quarks or gluons $f$ in the
channel $F$. The differential distribution of the antiprotons generated
by the hadronization of quarks (with the exception of the top quark)
and of gluons is denoted by $dN^{f}_{\bar{p}} / dE_{\bar{p}}$ and
depends on the nature of the species $f$. Of the various quantities
that are present in Eq.(\ref{susy_source}), $\langle\sigma v\rangle$
and $f(E_{\bar{p}})$ depend on the neutralino properties.
The antiproton production rate also depends on the distribution
${\rho}_{\chi}$ of the dark matter particles inside the galactic
halo. At this stage, some comments are in order~:

\noindent (i) The neutralino annihilation cross section
$\langle\sigma v \rangle$ is evaluated as in Ref. 
\cite{antip_noi}, with the parameters of the minimal supersymmetric standard 
model (MSSM) fixed at the following values: the lightest Higgs mass is
$m_h = 55$ GeV while the sfermion mass is $\tilde{m} = 500$ GeV
and $\tan \beta = 8$. The $\chi$ composition parameter
$P = a_1^2+a_2^2$ is fixed, for the three different cases
considered here, at the values $P=0.01$ (higgsino),
$P=0.5$ (mixture) and $P=0.99$ (gaugino).

\noindent (ii) For the $\bar{p}$ differential distribution
$f(E_{\bar{p}})$, we have evaluated the branching ratios
$B^{(F)}_{f}$ for all annihilation final states that may produce
antiprotons, i.e. direct production of quarks and gluons,
generation of quarks through the intermediate production of
Higgs bosons, gauge bosons and the top quark. The distributions
$dN^{f}_{\bar{p}}/dE_{\bar{p}}$ from the hadronization
of quarks (with the exception of the top quark) and gluons have been
computed by using the Monte Carlo code JETSET 7.2 \cite{BENGTSSON}.

\noindent (iii) The neutralino halo distribution
${\rho}_{\chi}$ is taken to be spherically symmetric. In the
axisymmetric coordinate system $r$ and $z$, the density profile is given by
\beq
\rho_{\chi} (r,z) \; = \; \rho_{\chi}(\odot) \;
\left\{ \frac {a^2 + r_\odot^2}{a^2 + r^2 + z^2} \right\} \;\; ,
\label{densite_neutralino}
\eeq
where $a = 3.5$ kpc is the core radius of the dark matter halo.
Particular care must be taken about the local neutralino density
$\rho_{\chi}(\odot)$, which depends on the LSP properties.
Actually, if the big-bang relic density $\Omega_{\chi} h^{2}$,
which we evaluate following \cite{omega_noi}, is too small to
account for the cosmological dark matter, the density of neutralinos
in the galactic halo should be corrected by a factor of
$\eta_{\chi}$. The latter deals with the fact that the neutralino density
is less than the halo density whenever $\Omega_{\chi} h^{2}$
is smaller than a minimal value of, say,
$(\Omega h^2)_{min} = 0.03$, which is compatible with the
observed rotation curve of the Galaxy. At cosmological distances,
this ratio is given by
\beq
\eta_{\chi} \; = \;
{\displaystyle \frac{\Omega_{\chi} h^{2}}{(\Omega h^2)_{min}}} \;\; ,
\eeq
whereas in the galactic halo, it may well be significantly larger.
For instance, there may be segregation between neutralinos and dark
baryons as a result of the dissipation of the latter. Therefore, we must
keep it in mind that $\eta_{\chi}$ could be as small as its cosmological
average, but it may also be way larger.

\section{ \bf The diffusion model for antiprotons }

Parker has studied the propagation of cosmic rays inside the Galaxy as
a consequence of their scattering by the irregularities of magnetic fields.
The presence of the latter is now firmly established by synchrotron
radiation far above the plane of our Galaxy as mentioned by Badwar
\cite{BADWAR}. Magnetic fields are also detected in other galaxies
\cite{SOFUE}. So the cosmic ray transport in the Galaxy
crucially depends on the diffusion across magnetic fields. In the following,
we will assume an isotropic diffusion with an empirical value for
the diffusion coefficient. Another proof of the existence of a diffusion
region in our Galaxy is provided by the recent diffuse $\gamma$-ray
observations, which point toward the presence of cosmic rays far
above the galactic disk.
As regards convection, it seems probable that cosmic rays as well
as the stellar wind from disk stars contribute to push out the magnetic
fields so that the region where the particles diffuse inflates at a speed
of the order of a few kilometers per second 
\cite{BADWAR,BLOEMEN_CR}. Because convection has been shown
to be negligible \cite{WLG}, we will disregard it and will focus our analysis
on the pure diffusion case. Thus, our Galaxy can be reasonably modelled
by a thin disk of atomic and molecular hydrogen with
$0 \leq r \leq R \; = \; 20$ kpc and $ |z| \leq h \; = \; 100$ pc, associated to
an extended region of diffusion containing irregular magnetic fields with
the same radial extension and $|z| \leq L$ = 3 kpc. Those various regions
are superimposed to the spheroidal halo of dark matter whose density
profile has already been given in relation (\ref{densite_neutralino}).
That two-zone diffusion model is in good agreement with the observed
primary and secondary nuclei abundances \cite{WLG}. Since their
discovery by Golden in 1979 \cite{GOLDEN}, cosmic antiprotons have been
thoroughly studied in the framework of the leaky box model
\cite{GAISSER}. We analyse here their propagation throughout the Galaxy
in the light of a two-zone diffusion model.

In a stationary regime, the propagation equation of cosmic antiprotons
may be expressed as
\beq
{\displaystyle \frac{\partial n_{\bar{p}}}{\partial t}} \; = \; 0 \; = \;
K \Delta n_{\bar{p}} \; - \; \Gamma_{\bar{p}} \, n_{\bar{p}} \; + \;
q_{\bar{p}} \;\; ,
\label{equation_de_propagation}
\eeq
where $n_{\bar{p}}(\EPB,r,z)$ is the density of antiprotons with
energy $\EPB$ at location $(r,z)$. In the right-hand side of relation
(\ref{equation_de_propagation}), the first term describes the diffusion
of antiprotons. The diffusion coefficient $K$ is constant at low energies,
but raises with the rigidity $p$ of the particle beyond a rigidity of
3 GV. This behaviour can be modelled by the form
\beq
K \; = \; 6\times 10^{23} \; \mbox{m\puis{2}} \cdot \mbox{s\puis{-1}}
\left(1 + \frac{p}{\mbox{3 GV}} \right)^{0.6} \;\; .
\label{diffusion_coefficient}
\eeq
The second term in Eq.(\ref{equation_de_propagation}) describes
the destruction of antiprotons by their interactions with the interstellar
medium. That destruction rate is shown to be very small but has not
been neglected in what follows. The collision rate of antiprotons
with the interstellar hydrogen is given by
\beq
\Gamma_{\bar{p}} \; = \; \sigma_{\bar{p} \, H} \; v_{\bar{p}} \; n_{H}
\;\; ,
\label{pbar_collision}
\eeq
where $\sigma_{\bar{p} \, H}$ is the total antiproton interaction
cross section with protons, $v_{\bar{p}}$ denotes the velocity and
$n_{H} = 1$ cm\puis{-3} is the average hydrogen density in the
thin matter disk. The last term in relation (\ref{equation_de_propagation})
deals with the various sources of antiprotons. Those species are produced by
the spallation of cosmic protons on the interstellar matter of the disk.
The antiproton production rate involves a convolution over the
incident cosmic proton energy spectrum $dn_{p}/d\EP$
\beq
q_{\bar{p}}^{\rm st}(\EPB) \; = \;
{\displaystyle \int_{\EPB}^{+ \infty}} \; d\EP \;
\left\{ \frac{dn_{p}}{d\EP}(\EP) \right\} \; \Gamma_{p}(\EP) \;
\YIELD \;\; .
\eeq
The collision rate $\Gamma_{p}$ of protons with the interstellar gas
is defined in just the same way as the collision rate $\Gamma_{\bar{p}}$
of antiprotons in relation (\ref{pbar_collision}). The antiproton differential
spectrum $Y_{\bar{p}}$ produced during a single proton spallation
was previously defined in Eq.(\ref{yield}). The supersymmetric
source term (\ref{susy_source}) may also be present if antiprotons are
produced by the mutual annihilation of neutralinos. This leads
to an additional antiproton flux.

We have solved Eq.(\ref{equation_de_propagation}) following
the analysis by Webber, Lee and Gupta \cite{WLG}. At the edge of
the domain where cosmic rays are confined, the particles escape freely
and the diffusion becomes inefficient. Thus the density vanishes at the
boundaries of the domain where cosmic rays are confined by diffusion.
This provides the initial conditions for solving the diffusion
equation. Then, because the problem has a cylindrical symmetry, the
densities $n_{p}$ and $n_{\bar{p}}$ may be expanded as series of the
Bessel functions of zeroth order
$J_{0} \left( \zeta_i x \right)$ where $\zeta_i$ is the ith zero
of $J_0$ and where $x = r/R$. Details may be found in \cite{WLG}.
The cosmic ray sources are located in the thin gaseous disk and we
have taken the radial distribution of supernovae remnants and pulsars
measured by Lyne {\it et al.} \cite{LMT}. As regards the spallation
mechanism, suffice it to say that an effective antiproton multiplicity
may be defined as
\beq
Y_{\bar{p}}^{\mbox{\scriptsize eff}} (\EPB) \; = \;
{\displaystyle \int_{\EPB}^{+ \infty}} \; d\EP \;
\left\{ \frac{\Phi_p(\EP)}{\Phi_p(\EPB)} \right\} \;
\YIELD \;\; ,
\eeq
so that no convolution with the cosmic proton energy spectrum
is needed any longer. The differential flux of protons of energy $\EP$
is denoted here by $\Phi_p(\EP)$.

In the case of the supersymmetric antiprotons, the resolution of
Eq.(\ref{equation_de_propagation}) is more involved.
The supersymmetric source term $q_{\bar{p}}^{\rm susy}$
now depends on both the galactocentric radius $r$ and on
the vertical coordinate $z$. In the solar neighbourhood, the
antiproton energy spectrum due to the mutual annihilations of
exotic particles is found to be~:
\beq
{\displaystyle \frac{d n_{\bar{p}}}{d \EPB}} (\odot) \; = \;
{\displaystyle \sum_{i=1}^{\infty}} \;
\left\{ \frac{4}{A_{i}} \right\} \;
\left\{ J_0 \left( \zeta_i \frac{R_\odot}{R} \right) \right\}
\; \left\{ J_1^{-2}(\zeta_i) \right\} \;
\left\{ \int_{0}^{L} \; {\cal F}_{i}(z) \; Q_{i}^{\rm susy}(z) \; dz \right\}
\;\; ,
\label{susy_pbar}
\eeq
where the Bessel transform $Q_{i}^{\rm susy}(z)$ is defined as
\beq
Q_{i}^{\rm susy}(z) \; = \; \int_{0}^{1} \; x \; dx \;
q_{\bar{p}}^{\rm susy}(E_{\bar{p}} , r=xR , z)
J_0 \left(\zeta_i \frac{R_\odot}{R} \right)\;\; .
\eeq
The vertical distribution ${\cal F}_{i}(z)$ is given by
\beq
{\cal F}_{i}(z) \; = \; \frac
{\sh \left\{ {\displaystyle \frac{S_i}{2}} (L - z) \right\}}
{\sh \left\{ {\displaystyle \frac{S_i}{2}} L \right\}} \;\; ,
\eeq
where the parameter $S_i$ is equal to $2 \zeta_i / R$. Finally,
the coefficient $A_i$ in relation (\ref{susy_pbar})
stands for
\beq
A_i \; = \; 2 h \Gamma_{\bar{p}} \; + \;
K S_i \coth \left\{ {\displaystyle \frac{S_i}{2}} L \right\} \;\; ,
\eeq
and only depends on the antiproton energy $\EPB$ through
the diffusion coefficient $K$ (\ref{diffusion_coefficient}) and
the collision rate $\Gamma_{\bar{p}}$ (\ref{pbar_collision}).
Expression (\ref{susy_pbar}) involves an integral on both the
galactocentric radius $r$ and on the height $z$. That integration
has been performed here with the specific form
(\ref{susy_source}) for the supersymmetric production rate
$q_{\bar{p}}^{\rm susy}$.

\section{Results and discussion}

We first show, in fig.~1, the standard interstellar $\bar{p}/p$ ratio
obtained in the framework of our diffusion model, as a function of
energy. As there is some uncertainty in the measurement of the spectral
index $\alpha$ of the high-energy proton spectrum, we considered the cases
$\alpha = 2.65$ (dashed line), 2.70 (intermediate solid curve) and 2.75
(dotted line). The $\bar{p}/p$ ratio decreases at low energy because the
low-energy antiprotons must be produced with a large backward momentum
in the centre-of-mass reference frame, and so their progenitors are
very high energy protons, whose density is very low. The harder the spectrum,
the larger the $\bar{p}/p$ ratio. Our results are in fair agreement with
the estimates of Gaisser and Schaefer \cite{GAISSER} collected in their
fig.~5. Differences arise from the improved treatment of the diffusion
mechanism with respect to the phenomenological leaky box model.

The antiproton signature $\bar{p}/p$ of neutralino pair-annihilations
in the halo must be compared with the standard $\bar{p}/p$ ratio discussed
above. Figure~2 features this ratio for three species of neutralinos, as
discussed in section~2. At low energy, the SUSY antiproton production
may exceed the standard signal in the pure gaugino and higgsino cases.
In the intermediate situation, it is well below the secondary antiproton flux
because the neutralino relic density is
$\Omega_{\chi} h^{2} \simeq 3 \times 10^{-3}$, only a tenth of
the minimal value $(\Omega h^2)_{min}$. This is not the case for
a pure gaugino or higgsino. Note that the suppression factor $\eta_{\chi}$
may be larger than 1/10 so that the dotted curve may well be shifted upward.

In the leaky box model, cosmic rays are distributed homogeneously
over the entire diffusion domain where they are confined by the
magnetic fields of the Galaxy. Their diffusion may be described from
a phenomenological point of view by the confinement time $\tau$,
which measures the rapidity with which the particles manage to escape
in the intergalactic medium. That confinement duration $\tau$ is
related to the column density, i.e. the amount of matter which the
cosmic rays cross during their erratic journey. Typical values of a few
g$\cdot$cm\puis{-2} are necessary to convert primary nuclei such as
carbon, oxygen or nitrogen into secondaries such as boron or beryllium.
This implies a confinement time in the matter disk of the Galaxy of order
$10^{7}$ years. Because the unstable isotope $^{10}$Be decays with a
half-life of 1.6 megayears (My), it provides a unique chronometer of
the time actually spent by cosmic rays since their production.
Measurements of its abundance relative to its stable partner indicate that
beryllium is confined during $\sim 10^{8}$ years, hence the need
of an extended region of diffusion which encapsulates the thin gaseous
disk. In the leaky box model, the supersymmetric production rate
$q_{\bar{p}}^{\rm susy}$ is averaged over the confinement region
and multiplied by the time $\tau_{g}$ it takes for a high-energy particle
to escape from that diffusion domain
\beq
{\displaystyle \frac{d n_{\bar{p}}}{d \EPB}} \; = \;
\langle \sigma v \rangle f(E_{\bar p}) \;
\left\{
{\displaystyle \frac{\langle \rho_{\chi}^{2} \rangle}{m_{\chi}^{2}}}
\right\} \; \tau_{g} \;\; .
\label{pbar_susy_lb}
\eeq
The square of the neutralino density is easily averaged over the diffusion
box. More uncertain is the determination of the confinement time
of antiprotons. Previous calculations \cite{antip_other,antip_noi}
are based on a typical escape time $\tau_{g}$ of order $10^{8}$ years.
The authors nevertheless note that estimates for that value may well
vary by an order of magnitude. Our two-zone diffusion model directly
incorporates that escape time, which depends on the cosmic ray distribution
throughout the Galaxy and on the energy through the diffusion coefficient.
The escape time from the disk $\tau_{d}$ and from the whole Galaxy $\tau_{g}$
may also be derived from that model. For protons or antiprotons, the net
production rate $Q$ in the entire Galaxy may be computed. A fraction of the
particles that are produced are destroyed by spallation on the interstellar gas
of the disk. What remains is the escape rate $\dot{N}_{\rm esc}$, provided
that steady state is achieved
\beq
Q \; = \; \Gamma N_{d} + \dot{N}_{\rm esc} \;\; .
\eeq
The confinement time of cosmic rays in the thin gaseous disk and in the
whole domain of diffusion are related to the escape rate through
\beq
\dot{N}_{\rm esc} \; = \;
{\displaystyle \frac{N_{d}}{\tau_{d}}} \; = \;
{\displaystyle \frac{N_{g}}{\tau_{g}}} \;\; ,
\eeq
where the total number of particles in these two regions are respectively 
denoted by $N_{d}$ and $N_{g}$. For both protons and antiprotons, we have 
evaluated the various escape times in the framework of our diffusion model. We 
find that they are fairly similar.  The solid curve
in fig.~3 features the variations of the antiproton escape time $\tau_{g}$
as a function of kinetic energy. The dashed line stands for the escape time from
the matter disk. Both escape times decrease with energy. As a matter of fact,
the diffusion coefficient increases at high rigidity so that particles diffuse, and
therefore escape, more easily. From 100 MeV to 1 GeV, the disk confinement
duration is $\tau_{d} \sim$ 11 My while the escape time from the Galaxy as a
whole is $\tau_{g} \sim$ 180 My, a factor of 2 above the value currently
used in the previous estimates of the antiproton flux. In fig.~4, we focus on
the antiproton signal of supersymmetric dark matter. For purposes of comparison,
the ratio of the leaky box prediction (\ref{pbar_susy_lb}) to the exact result
(\ref{susy_pbar}) is plotted as a function of the cosmic ray kinetic energy.
The solid line corresponds to an escape time $\tau_{g}$ of
$3 \times 10^{15}$~s, independent of the energy. At low energy, the leaky
box model underestimates the antiproton flux by a factor of 2. On the contrary,
it is too optimistic at high energy by a factor $\sim$ 3.
If in relation (\ref{pbar_susy_lb}), we use the value of the escape time 
$\tau_{g}$, which has been derived in fig.~3, the situation considerably improves
as featured by the dashed curve. The leaky box model may actually be used,
provided the escape times are correctly derived with a diffusion model. The
excess over relation (\ref{susy_pbar}) is at most 30\% for a kinetic energy
of 1 GeV.

In this letter, we have used a two-zone diffusion model to describe the
propagation of antiprotons throughout the Galaxy. We have estimated
the conventional and supersymmetric antiproton yields on the Earth.
As regards the neutralino pair-annihilations in the halo, a correct
treatment of diffusion improves the old leaky box estimates. We have
presented here three characteristic examples of neutralino composition.
A more exhaustive investigation of the supersymmetric parameter space
is now necessary.

\vskip 0.5cm

{\bf Acknowledgements} \\ This work has been carried out under the auspices
of the Human Capital and Mobility Programme of the European Economic
Community, under contract number CHRX-CT93-0120 (DG 12 COMA).
We acknowledge the financial support of a Collaborative
Research Grant from NATO under contract CRG 930695. R.~Taillet
would like to thank the Institute for Nuclear Physics and Astrophysics
(INPA) of the Lawrence Berkeley Laboratory for its kind hospitality
and financial support. R.~Taillet initiated this work during his stay at
INPA in the summer of 1995. Also a fellowship of the Universit\`a di Torino
is gratefully acknowledged by G. Mignola.

\newpage
{\large\bf Figure Captions}
\vskip 1.cm

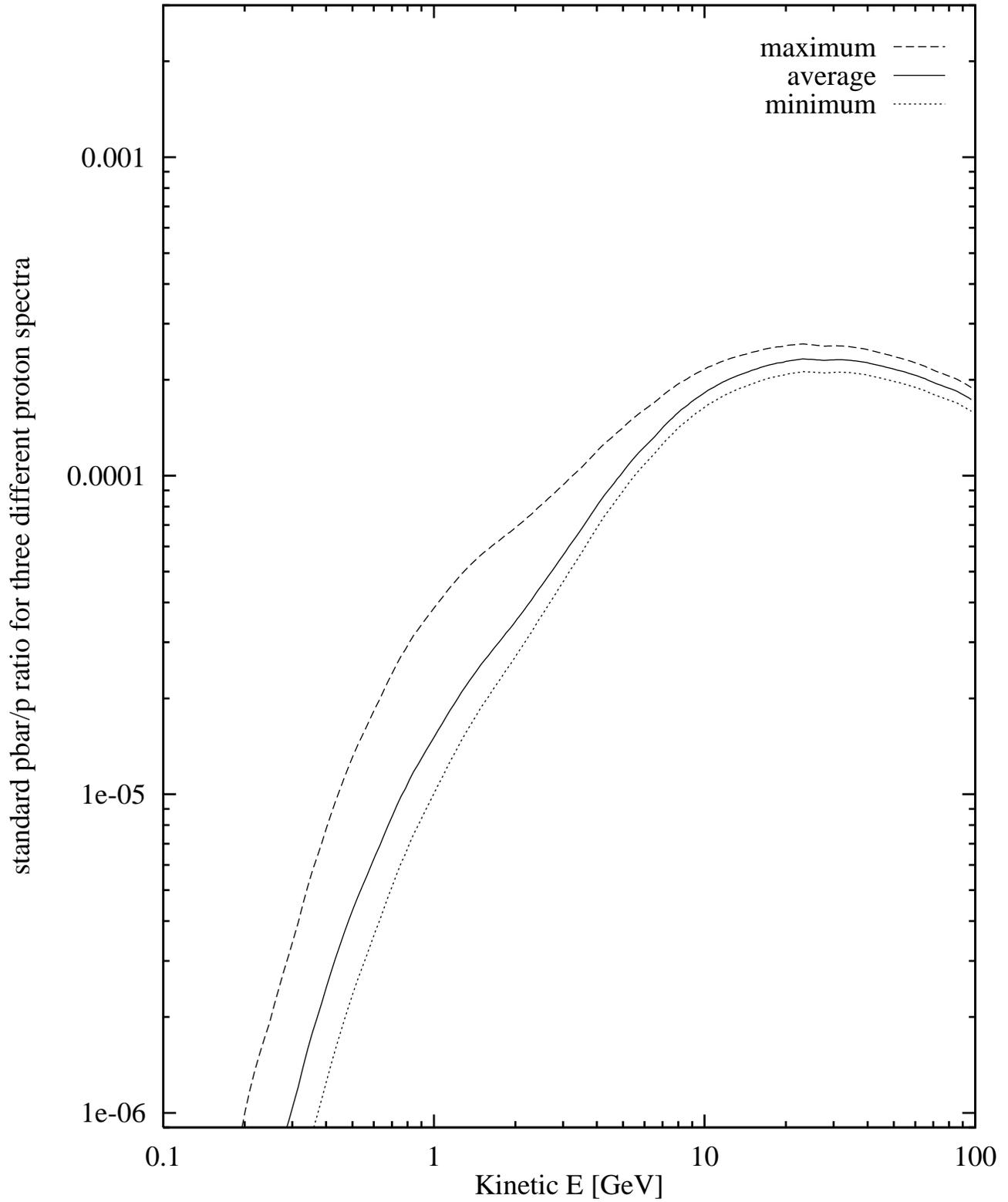
Fig~1 : The $\bar{p}/p$ ratio obtained in the framework
of a diffusion model is displayed as a function of energy.
The sources of antiprotons are the spallation reactions of
cosmic ray primaries on the interstellar gas. The spectral
index of the proton flux has been varied from 2.65 (top curve)
to 2.75 (bottom curve).

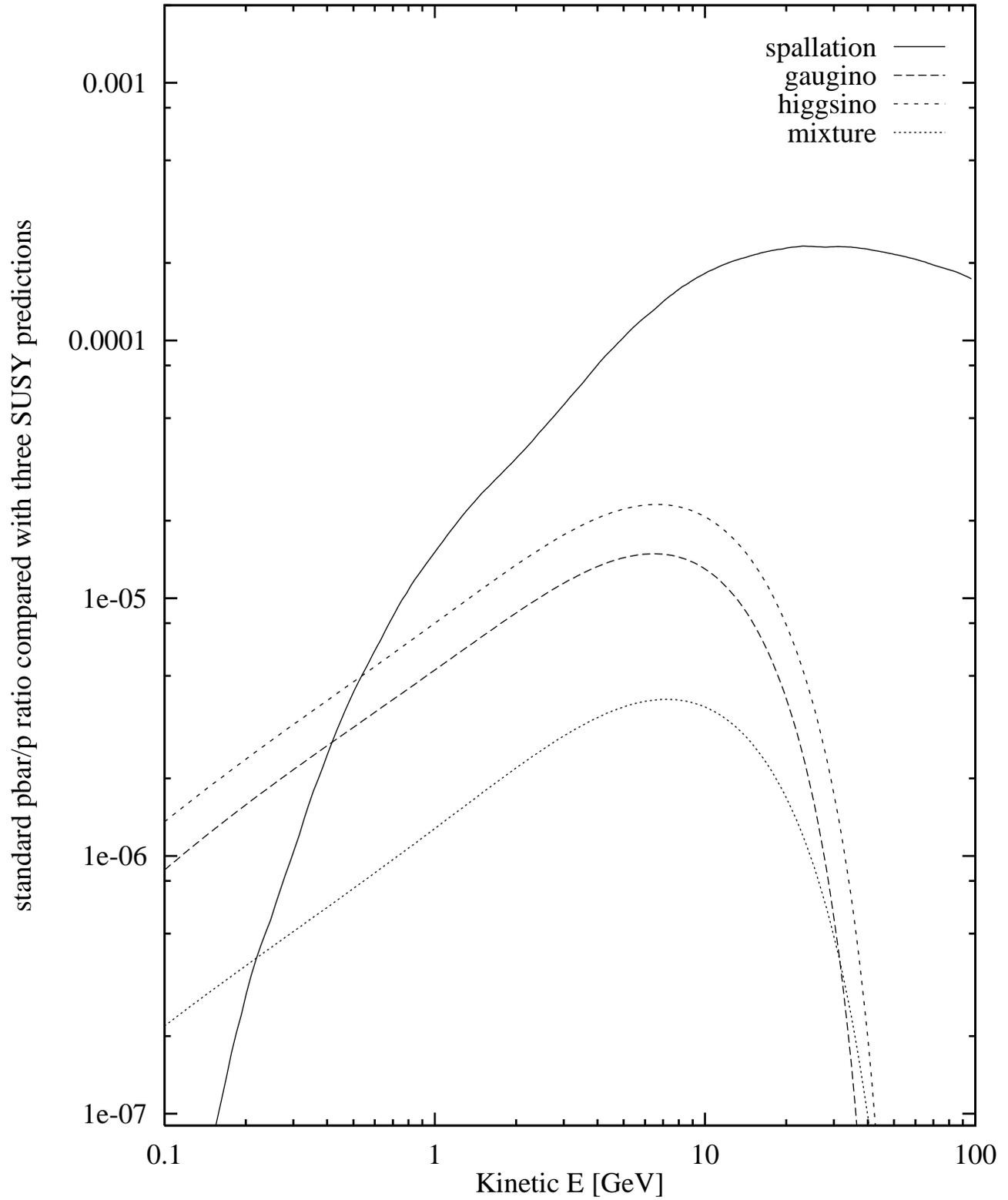
Fig~2 : The $\bar{p}/p$ ratio is featured as a function of energy,
for various antiproton sources: standard spallations (solid line)
assuming a spectral index of 2.7 for the primary proton distribution,
and neutralino pair annihilations, for three different neutralino
compositions (gaugino, higgsino, and a mixture).

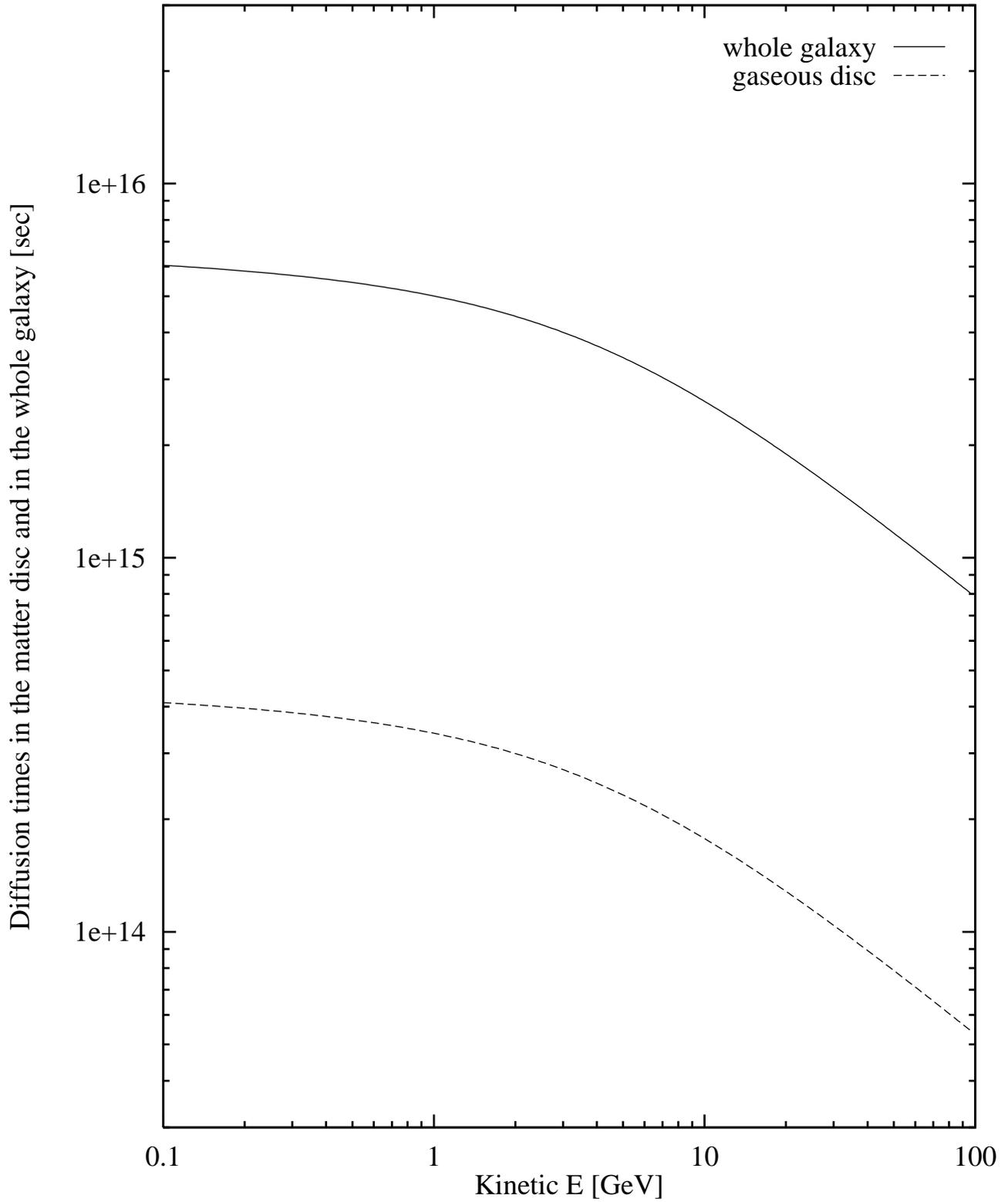
Fig~3 : The diffusion time scales of antiprotons in the thin gaseous disk
(dashed line) and in the whole Galaxy (solid line) are plotted as a function
of energy. Those time scales are inferred in the framework of the diffusion
model discussed in the text.

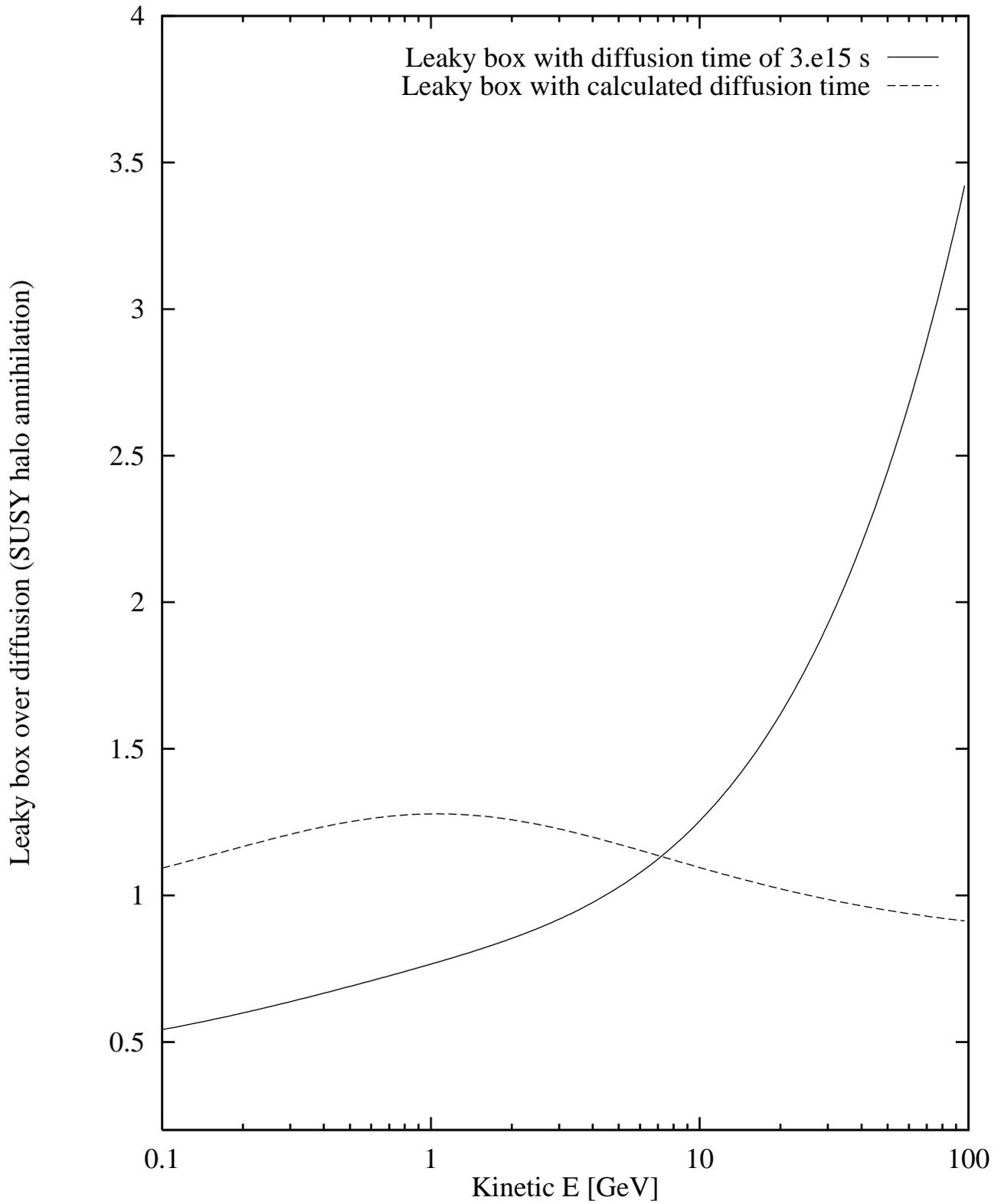
Fig~4 : The ratio of leaky box results over direct diffusion estimates
are presented as a function of energy, for the antiprotons
from neutralino pair annihilations. The solid line is obtained by
setting in the leaky box model the antiproton escape time scale at
the constant value of $3\times 10^{15}$ s. In the case featured by the
dashed curve, this escape time scale has been borrowed from fig.~3.


\newpage
\begin{thebibliography}{99}

\bibitem{AMS} S.~Ahlen {\it et al.}, {\sl Nucl. Instrum. Meth.}
{\bf A350} (1994) 351.

\bibitem{BLOEMEN_GAMMA}
H.~Bloemen, {\sl ARA\&A}~{\bf 27} (1989) 469.

\bibitem{GRO_CEPHEUS}
Digel S.~W., Grenier I.~A., Heithausen A.,
Hunter S.~D., Thaddeus P., 1994, {\sl BAAS}~{\bf 185},
number 120.02.

\bibitem{BENGTSSON} H.~U.~Bengtsson and T.~Sj\"ostrand,
{\sl Comput. Phys. Commun.}~{\bf 46} (1987) 43;\\
T.~Sj\"ostrand, preprints CERN-TH.6488/92 and CERN-TH.7111/93.

\bibitem{EHNOS} J.~Ellis, J.~S.~Hagelin, D.~V.~Nanopoulos,
K.~Olive and M.~Srednicki, {\sl Nucl. Phys.}~{\bf B238} (1984) 453.

\bibitem{GKJ} See, for a review of the detection strategies:
G.~Jungman, M.~Kamionkowski and K.~Griest,
{\sl Phys. Rep.}~{\bf 267} (1996) 195, and references quoted therein.

\bibitem{th_dir}
J.~Ellis and R.~A.~Flores, {\sl Phys. Lett.}~{\bf B263} (1991) 259;
{\sl Nucl. Phys.}~{\bf B400} (1993) 25 and
{\sl Phys. Lett.}~{\bf B300} (1993) 175;\\
A.~Bottino, V.~de~Alfaro, N.~Fornengo, G.~Mignola and S.~Scopel,
{\sl Astropart. Phys.}~{\bf 2} (1994) 77;\\
C.~Bacci {\it et al.} (BRS Collaboration), {\sl Phys. Lett.}~{\bf B295} (1992) 330;\\
V.~Berezinsky, A.~Bottino, J.~Ellis, N.~Fornengo, G.~Mignola and S.~Scopel,
preprint CERN--TH 95--206, to appear in {\sl Astropart. Phys.}.

\bibitem{exp_dir} See, for instance,
R.~Bernabei, {\sl Riv. N. Cim.}~{\bf 18} (1995) N.5 and
L. Mosca, invited talk at TAUP 95 (Toledo, September 1995) to appear
in {\sl Nucl. Phys.}~{\bf B} (Proc. Suppl.).

\bibitem{nu}
G.~Gelmini, P.~Gondolo and E.~Roulet, {\sl Nucl. Phys.}~{\bf B351}
(1991) 623;\\
M.~Kamionkowski, {\sl Phys. Rev.}~{\bf D44} (1991) 3021;\\
F.~Halzen, T.~Stelzer and M.~Kamionkowski, {\sl Phys. Rev.}~{\bf D45}
(1992) 4439;\\
M.~Mori {\it et al.}, {\sl Phys. Rev.}~{\bf D48} (1993) 5505;\\
A.~Bottino, N.~Fornengo, G.~Mignola and L.~ÊMoscoso,
{\sl Astropart. Phys.}~{\bf 3} (1995) 65;\\
V.~Berezinsky, A.~Bottino, J.~Ellis, N.~Fornengo, G.~Mignola and S.~Scopel,
preprint CERN--TH 96--42.

\bibitem{antip_other}
J.~Ellis, R.~A.~Flores, K.~Freese, S.~Ritz, D.~Seckel and J.~Silk,
{\sl Phys. Lett.}~{\bf B214} (1988) 403;\\
A.~Bouquet, P.~Salati and J.~Silk, {\sl Phys. Rev.}~{\bf D40} (1989) 3168;\\
H.~U.~Bengtsson, P.~Salati and J.~Silk, {\sl Nucl. Phys.}~{\bf B346}
(1990) 129;\\
M.~Kamionkowski and M.~Turner, {\sl Phys. Rev.}~{\bf D43} (1991) 1774;\\
V.~Berezinsky, A.~Bottino and G.~Mignola, {\sl Phys. Lett.}~{\bf B325}
(1994) 136;\\
L.~Bergstr{\"o}m and J.~Kaplan, {\sl Astropart. Phys.}~{\bf 2} (1994) 261;\\
G.~Jungman and M.~Kamionkowski, {\sl Phys. Rev.}~{\bf D49} (1994) 2316
and {\bf D51} (1995) 3121.

\bibitem{antip_noi}
A.~Bottino, C.~Favero, N.~Fornengo and G.~Mignola,
{\sl Astropart. Phys.}~{\bf 3} (1995) 77.

\bibitem{omega_noi} See for instance A.~Bottino, V.~de~Alfaro, N.~Fornengo,
G.~Mignola and M.~Pignone, {\sl Astropart. Phys.}~{\bf 2} (1994) 67.

\bibitem{BADWAR} G.~D.~Badwar and S.~A.~Stephens,
{\sl Ap.J.}~{\bf 212} (1977) 494.

\bibitem{SOFUE} Y.~Sofue, M.~Fujimoto and R.~Wielebinski,
{\sl ARA\&A}~{\bf24} (1986) 459.

\bibitem{BLOEMEN_CR} J.~B.~G.~M.~Bloemen,
{\sl Astron. Astrophys.}~{\bf 267} (1993) 372.

\bibitem{WLG} W.~R.~Webber, M.~A.~Lee and M.~Gupta,
{\sl Ap.J.}~{\bf 390} (1992) 96.

\bibitem{GOLDEN} R.~L.~Golden, S.~Horan and B.~G.~Mauger,
{\sl Phys. Rev. Lett. }~{\bf 43} (1979) 1196.

\bibitem{GAISSER} T.~K.~Gaisser and R.~K.~Schaefer,
{\sl Ap.J.}~{\bf 394} (1992) 174.

\bibitem{LMT} A.~G.~Lyne, R.~N.~Manchester and J.~H.~Taylor,
{\sl MNRAS}~{\bf 213} (1985) 613.










\end{thebibliography}
\end{document}